\documentclass[graybox]{svmult}

\usepackage{mathptmx}       
\usepackage{helvet}         
\usepackage{courier}        
\usepackage{type1cm}        
\usepackage{makeidx}         
\usepackage{graphicx}        
\usepackage{multicol}        
\usepackage[bottom]{footmisc}

\makeindex             

\newcommand{\HI}{{\sc{HI}}}
\newcommand{\msun}{$M_\odot$}

\begin{document}

\title*{Measuring the halo mass function in loose groups}
\author{D.J. Pisano, D.G. Barnes, B.K. Gibson, L. Staveley-Smith, K.C. Freeman, and V.A. Kilborn}
\authorrunning{Pisano et al.}
\institute{D.J. Pisano \at West Virginia University Dept. of Physics, P.O. Box 6315, Morgantown, WV 26506, USA , \email{djpisano@mail.wvu.edu}
\and D.G. Barnes \& V.A. Kilborn \at Centre for Astrophysics and Supercomputing, Swinburne University, Hawthorn, VIC 3122, Australia
\and B.K. Gibson \at  Centre for Astrophysics, University of Central Lancashire, Preston, PR1 @HU, UK
\and L. Staveley-Smith \at School of Physics, M013, University of Western Australia, Crawley, WA 6009, Australia
\and K.C. Freeman \at RSAA, Mount Stromlo Observatory, Cotter Road, Weston, ACT 2611, Australia}
%
%
\maketitle

\vskip-1.2truein

\abstract{Using data from our Parkes \& ATCA \HI\ survey of six groups analogous to the Local Group, we find that the \HI\ mass
function and velocity distribution function for loose groups are the same as those for the Local Group.  Both mass functions confirm that the ``missing satellite" problem exists in other galaxy groups.}

\section{Project Overview}
\label{sec:1}

Cold dark matter (CDM) models of galaxy formation predict that the Local Group should contain about 300 dark matter halos but there is an
order of magnitude fewer galaxies observed \cite{klypin99,moore99}.  While the "missing satellite" problem can be mitigated by the inclusion of 
baryon physics in CDM models or alternate forms of dark matter, it is important to establish how this problem is manifest beyond the Local Group.

We have conducted a \HI\ survey of six loose groups of galaxies that are analogous to the Local Group.  The six groups are composed of 
only spiral and irregular galaxies that have mean separations of $\sim$550\,kpc.  The groups have average diameters of 1.6\,Mpc and 
have M$_{virial}\sim$10$^{11.7-13.6}$\,\msun; they are similar to the Local Group in all these ways.  
Details on our observations, data reduction, and our 
search for \HI\ clouds in the groups can be found in \cite{pisano07}.  The survey identified a total of 63 group galaxies with all of the new
detections having properties consistent with being typical dwarf irregular galaxies.  

\section{Halo Mass Functions}
\label{sec:2}

\begin{figure}[t]
\includegraphics[width=0.5\textwidth]{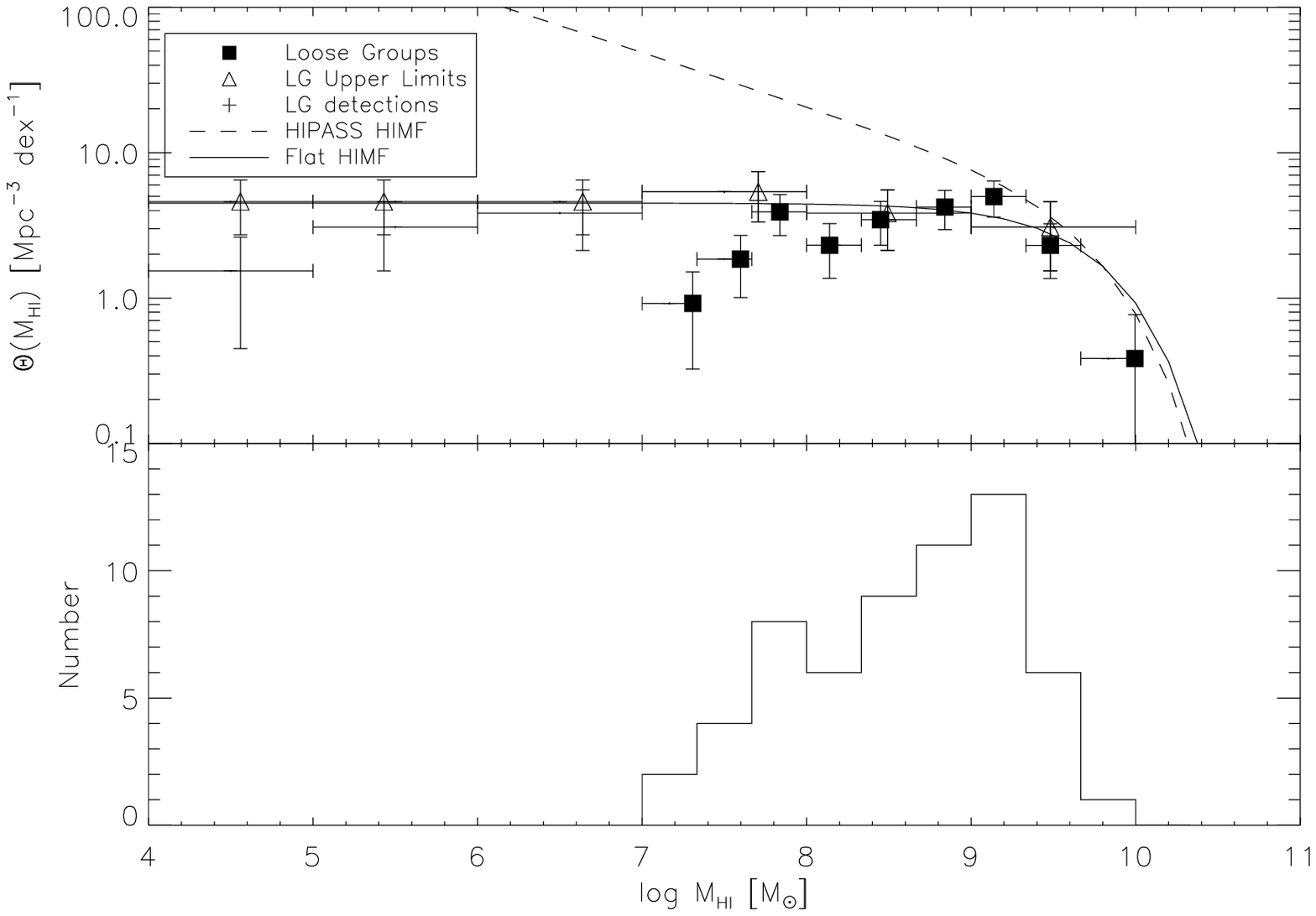}\hfill
\includegraphics[width=0.5\textwidth]{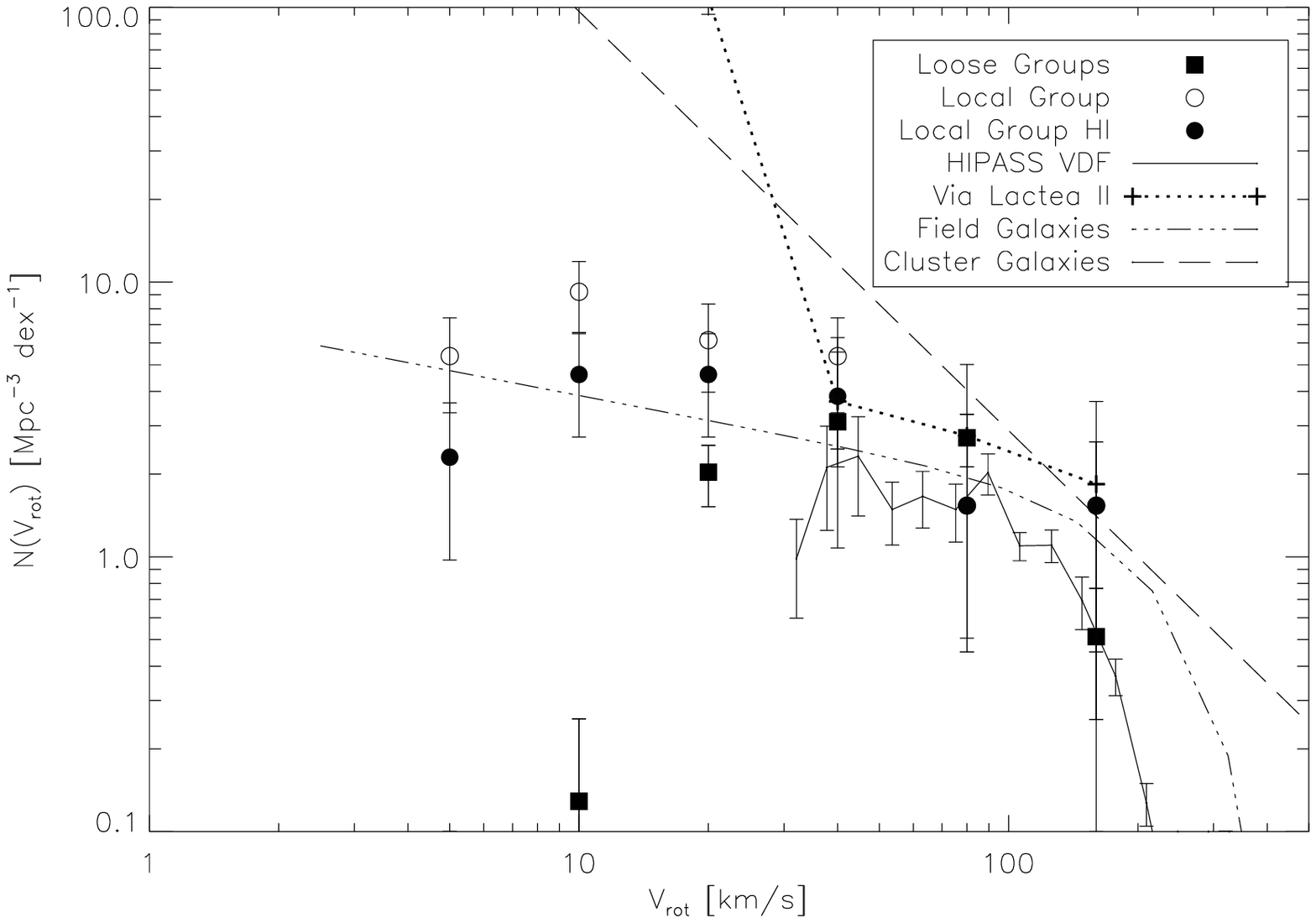}
\caption{Left:  The \HI MF for loose groups as compared to that for the Local Group galaxies with \HI\ detections and Local Group
galaxies with \HI\ detections and upper limits.  Also shown is the \HI MF from HIPASS \cite{zwaan05} and a flat \HI MF. Right:  The VDF 
 for the loose groups compared to all Local Group galaxies, only those detected in \HI, the HIPASS VDF from \cite{zwaan05},
field galaxies from \cite{gonzalez00}, cluster galaxies from \cite{desai04}, and the theoretical predictions from Via Lactea II \cite{diemand08}.  Aside from the loose and Local Group data, all other functions have been arbitrarily renormalized.}
\label{fig:pisano1}
\end{figure}

Using the survey completeness from \cite{pisano07} and our catalog of group galaxies, we constructed both a \HI\ mass function (\HI MF) and
a circular velocity distribution function (VDF) for the six loose groups as shown in Figure~\ref{fig:pisano1}.  The figure shows that both the 
\HI MF and VDF for the Local Group are not atypical, but are consistent with those for the six loose groups.  The \HI MF for low density regions,
such as loose groups (and including the Local Group), is consistent with being flatter than the \HI MF in the field as was found by \cite{zwaan05}.
The VDF for loose groups has a consistent low mass slope to field galaxies and HIPASS galaxies \cite{gonzalez00,zwaan10}, but is much
shallower than is predicted by simulations \cite{diemand08} or observed in galaxy clusters \cite{desai04}.  For a more complete discussion of 
these results, see Pisano et al. (2011, in preparation).  

%



\begin{thebibliography}{99.}
\bibitem{desai04} Desai, V., Dalcanton, J.~J., Mayer, L., Reed, D., Quinn, T., \& Governato, F.\ 2004, MNRAS, 351, 265 

\bibitem{diemand08} Diemand, J., Kuhlen, M., Madau, P., Zemp, M., Moore, B., Potter, D., \& Stadel, J.\ 2008, Nature, 454, 735 

\bibitem{gonzalez00} Gonzalez, A.~H., Williams, K.~A., Bullock, J.~S., Kolatt, T.~S., \& Primack, J.~R.\ 2000, ApJ, 528, 145 

\bibitem{klypin99} Klypin, A., Kravtsov, A.V., 
Valenzuela, O., \& Prada, F., 1999, ApJ, 522, 82

\bibitem{moore99} Moore, B., Ghigna, S., Governato, 
F., Lake, G., Quinn, T., Stadel, J., \& Tozzi, P., 1999, ApJ, 524, L19

\bibitem{pisano07} Pisano, D.~J., Barnes, 
D.~G., Gibson, B.~K., Staveley-Smith, L., Freeman, K.~C., \& Kilborn, 
V.~A.\ 2007, ApJ, 662, 959

\bibitem{zwaan05} Zwaan, M.A., Meyer, M.J., Staveley-Smith, L., \& Webster, R.L., 2005, MNRAS, 359, L30

\bibitem{zwaan10} Zwaan, M.A., Meyer, M.J., \& Staveley-Smith, L., 2010, MNRAS, 403, 1969

\end{thebibliography}
\end{document}